\documentclass[12pt,thmsa]{article}
\usepackage{amssymb}

\usepackage{sw20lart}

\usepackage{epsfig}
\newcounter{iiicounter}
\def\biilist{%
\begin{list}{{\hss\it\roman{iiicounter})\/}}%
{\usecounter{iiicounter}\setlength{\topsep}{0pt}%
\setlength{\itemsep}{0pt}%
\setlength{\leftmargin}{0.8cm}%
\setlength{\labelwidth}{0.65cm}%
\setlength{\labelsep}{0.15cm}}}
\def\eiilist{\end{list}}

\def\grad{\mathop{\bf grad}\nolimits}
\def\`#1{\if#1i{\accent18 \i}\else{\accent18 #1}\fi}
\def\'#1{\if#1i{\accent19 \i}\else{\accent19 #1}\fi}

\input tcilatex
\title{Geomagnetic Field and Air Shower Simulations.
}
\author{{\large\rm A.~Cillis and S.~J.~Sciutto
\vrule depth \baselineskip width 0pt}\\
{\large\it Departamento de F\'isica}\\
{\large\it Universidad Nacional de La Plata}\\
{\large\it C. C. 67 - 1900 La Plata}\\
{\large\it Argentina}}
\begin{document}
\baselineskip=17pt
\input pstricks
\maketitle
\vfil

\begin{abstract}
The influence of the geomagnetic field on the development of air
showers is studied.
The well known International Geomagnetic Reference Field was included
in the AIRES air
shower simulation program as an auxiliary tool to allow calculating very
accurate estimations of the geomagnetic field given the geographic
coordinates, altitude above sea level and date of a given event.
Some test simulations made for representative cases indicate that some
quantities like the lateral distribution of muons experiment significant
modifications when the geomagnetic field is taken into account.

\end{abstract}

\vfil\eject

\section{Introduction}

The charged particles of an air shower interact with the Earth's magnetic
field. One of the effects of such interaction is that of curving the
particles' paths.

In order to take into account such effect, we have incorporated the
geomagnetic field (GF) in the AIRES air shower simulation program \cite
{aires,sergio} allowing to simulate the showers in presence of the
field at any location and time.

We have first studied the principal characteristics of the GF: Its origin,
magnitude and variations. Thereafter we have analyzed the different GF
models that are normally used: The dipolar models \cite{campoB} which as their
name suggest, consider the GF as generated by magnetic dipoles whose
magnitude and orientation are set to fit the experimental measurements; the
so-called International Geomagnetic Reference Field (IGRF) \cite{IGRF}, a
more elaborated model, based on a high-order harmonic expansion whose
coefficients are fitted with data coming from a network of geomagnetic
observatories all around the world.

In this paper we present a simple comparative analysis of the different
models. One of the conclusions that comes out from our analysis is that the
dipolar models are not useful to evaluate the GF at a given arbitrary
location with enough accuracy, a more sophisticated model like a high order
series expansion is needed instead. On the other hand, the predictions of
the IGRF proved to match with the corresponding experimental data with
errors that are at most a few percent.

We have therefore selected the IGRF to link it to the simulation program
AIRES, as an adequate model to synthesize the GF at a given geographical
location and time.

This work is organized as follows. In section 2 we start describing the GF.
In sections 3 and 4 we mention some of the different models of the GF that
exist at present. We analyze them and in section 5 we compare the
predictions of these models with the experimental data.

The practical implementation of the GF in the AIRES program is
discussed in section 6. Section 7 is devoted to the analysis of the
effects of the GF on some air shower observables, and in section 8 we
place our final remarks and conclusions.

\section{Description of the geomagnetic field}

The Earth's magnetic field is normally described by seven parameters, namely,
declination (D), inclination (I), horizontal intensity (H), vertical
intensity (Z), total intensity (F) and the north (X) and east (Y) components
of the horizontal intensity. D is the angle between the horizontal component
of the magnetic field and the direction of the geographical north and I is
the angle between the horizontal plane and the total magnetic field. It is
considered positive when the magnetic field points downwards. Also Z is
positive when I is positive \cite{campoB}.

The geomagnetic poles are located, by definition, in the places where the
field lines are perpendicular to the Earth's surface. The physical
locations of the magnetic poles are actually areas rather than single
points. Because of the changing nature of the GF, the locations of the
magnetic poles also change. The current location of the N (S) magnetic pole
is approximately $78.5^\circ$ N and $103.4^\circ$ W ($65^\circ$ S and
$139^\circ$ W).

The GF is generated by internal and external sources. The first ones
are related
to processes in the interior of the Earth's core while the external
sources would
be related to ionized currents in the high atmosphere \cite{campoB}. The
experimental measurements show that the internal field is significantly
greater than the external contribution. The former goes from 20.000 to
70.000 nanoteslas (nT, 1 nT = $10^4$ gauss) while the external contribution is
around 100 nT.

The GF evolves with time. The rates of change of the different components
are not uniform over position and time and can be classified as
follows \cite{campoB}:
\begin{itemize}
\item \textbf{Secular variations:} Extended over years with generally
smooth increases or decreases in the field. They are originated by the
internal field and are the least understood of all the kinds of changes
that affect the GF. The values of the secular variation of the components of
the GF go from 10 nT per year up to 150 nT/year and up to 6 to 10 arc
minutes/year for D and I.

\item \textbf{Periodic variations:} They are originated by the external
field and in general amount to less than 100 nT. The characteristic periods
are 12 hs, 1 day, 27 days, 6 months and 1 year. They are related with the
Earth's rotation and with the Solar and Lunar influence.

\item \textbf{Magnetic storms: }Sudden disturbances in the GF which may
last from hours up to several days and rarely modify the field in more than
500 nT.
\end{itemize}

\section{Models of the geomagnetic field}

\subsection{Dipolar model}

The simplest way to model the GF is to assume that it is generated by a
magnetic dipole. Two alternative dipolar models are normally taken into
account \cite{campoB}:
\begin{itemize}
\item \textbf{Dipolar central model}: It is assumed that the field is
generated by a dipole with origin in the center of the Earth, inclined some
degrees with respect of the rotation axis. This model is used to define the
geomagnetic coordinates (they are the coordinates from the dipole's axis): a)
Geomagnetic latitude, $\varphi^{*}$, measured from the geomagnetic
equator, defined as the plane normal to the dipole's axis and passing
through the center of the Earth, b) Geomagnetic
longitude, $\lambda ^{*}$, measured eastwards from the meridian half-plane
containing the geographical south pole.

The geomagnetic coordinates $(\varphi ^{*},\lambda ^{*})$ of each
point P have a one-to-one relationship
with the point's geographical coordinates $(\varphi ,\lambda )$. The
transformation formulae can be deduced using spherical trigonometry or
rotation matrices.

\item \textbf{Dipolar eccentric model}: The field is generated by a
dipole displaced from the center of the Earth.
\end{itemize}

\subsection{Harmonic analysis of the geomagnetic field}

When a more accurate reproduction of the field is needed, it is necessary to
go beyond the dipole approximation and make a higher order harmonic analysis
of the GF \cite{campoB}:
The GF is modeled like the magnetostatic field whose sources are currents
located in the interior of the Earth. Then for points located near the
Earth's surface, it is possible to calculate the GF using the scalar
magnetic potential $\phi$, via $\mathbf{B}=-\grad\phi$. Such
scalar potential satisfies Laplace's equation, $\nabla ^2\phi =0$.

Due to the spherical symmetry of the problem, the solution can be
conveniently expressed in terms of Legendre functions. The scalar magnetic
field can be expanded in terms of the geographical coordinates as

\begin{equation}
\phi =a\dsum\limits_{n=1}^N\dsum\limits_{m=0}^n\QOVERD( ) {a}{r}^{n+1}\left[
g_{nm}\cos m\lambda +h_{nm}\sin m\lambda \right] P_n^m(\cos \varphi )
\label{soluc}
\end{equation}
where $a$ is the mean radius of the Earth (6371.2 km), $r$ is the radial
distance from the center of the Earth, $\lambda $ is the longitude
eastwards from
Greenwich, $\varphi $ is the geocentric colatitude, and $P_n^m(\cos \varphi )$
is the associated Legendre function of degree $n$ and order $m$, normalized
according to the convection of Schmidt. $N$ is the maximum spherical harmonic
degree of the expansion.

\section{International Geomagnetic Reference Field}

The International Geomagnetic Reference Field (IGRF) \cite{IGRF} is a
parameterization of experimental values using equation (\ref{soluc}). Sets of
spherical harmonic coefficients ($g_{nm}$ and $h_{nm})$ at 5-year intervals
starting from 1900 are evaluated. They are determined from the measurements
of the components of the field made at the Earth's surface (geomagnetic
observatories and satellite observations). Coefficients for dates between
5-year epochs are obtained by linear interpolation between the corresponding
coefficients for the neighboring epochs. At present, the model includes
secular variation terms for forward continuation of it for the years 1995 to
2000.

The error of the field components and the D and I angles are respectively
less than 500 nT and 30 arc minutes. These errors are relatively small (for
example they amount to a few percent in the case of the field intensity F)
and this makes the IGRF model a very useful tool to estimate the GF at any
geographic location and any time belonging to its validity interval.

\section{Analysis of the different models}

We have evaluated the results coming from the models already introduced in a
variety of situations, in order to establish which of them is the most
convenient to cover the needs arising in an air shower simulation algorithm.

\begin{figure}
\begin{center}
\epsfig{file=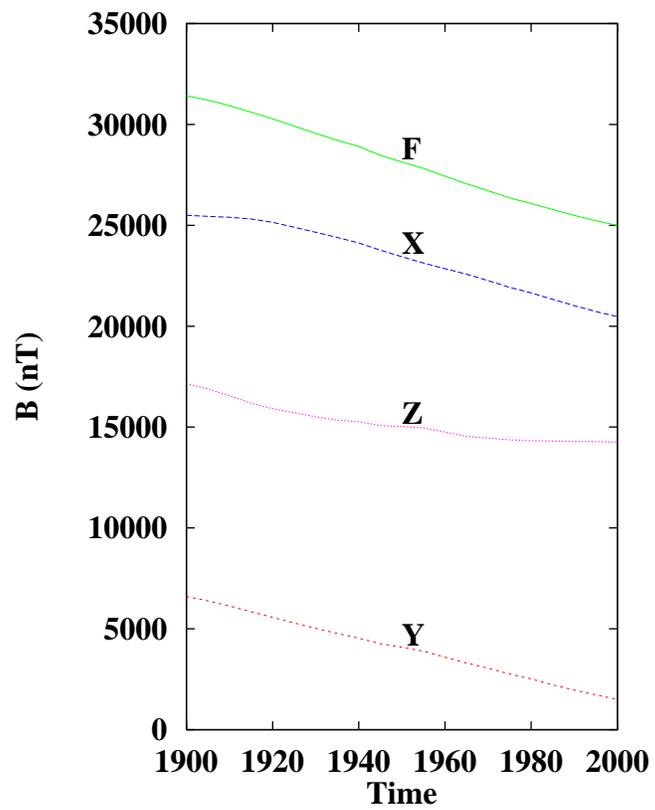}
\end{center}
\caption{Secular variation of the GF (1900-2000). Site: El Nihuil
(Argentina).}
\end{figure}
To start with, we have used the IGRF series expansion to study the secular
variation of the GF at a fixed site, namely the El Nihuil site
(lat.~$35.2^\circ$ S, long. $69.2^\circ$ W, altitude 1400 m.a.s.l.)
located in Argentina.
In figure 1 the components of the GF are displayed as functions of time for
the years 1900-2000. The rate of change of the field intensity F is roughly
20 \%/century. The variation of the field components with time must therefore
be taken into account if it is necessary to reproduce the GF within a few
percent error limit.

We have also studied the spatial variation of the GF at a given fixed time.
In figure 2 the components of the GF are plotted versus the geographic
latitude for the fixed longitude of $69^\circ$ W (longitude of the El Nihuil
site). There are two main conclusions that can be extracted from these
plots: (1) The spatial variations of the field components are very
important: Notice, for example, that the field intensity, F, goes from
a minimum of 24000 nT up
to 58000 nT, that is, more than twice the minimum value, (2) The predictions
of the dipole models, either centered or eccentric, can differ in more than
30 \% with respect to the IGRF and hence with experimental values. Therefore
these models cannot be used to produce safe estimations of the GF at any
given arbitrary location.
\begin{figure}
\begin{displaymath}
\begin{array}{cc}
\hbox{\epsfig{file=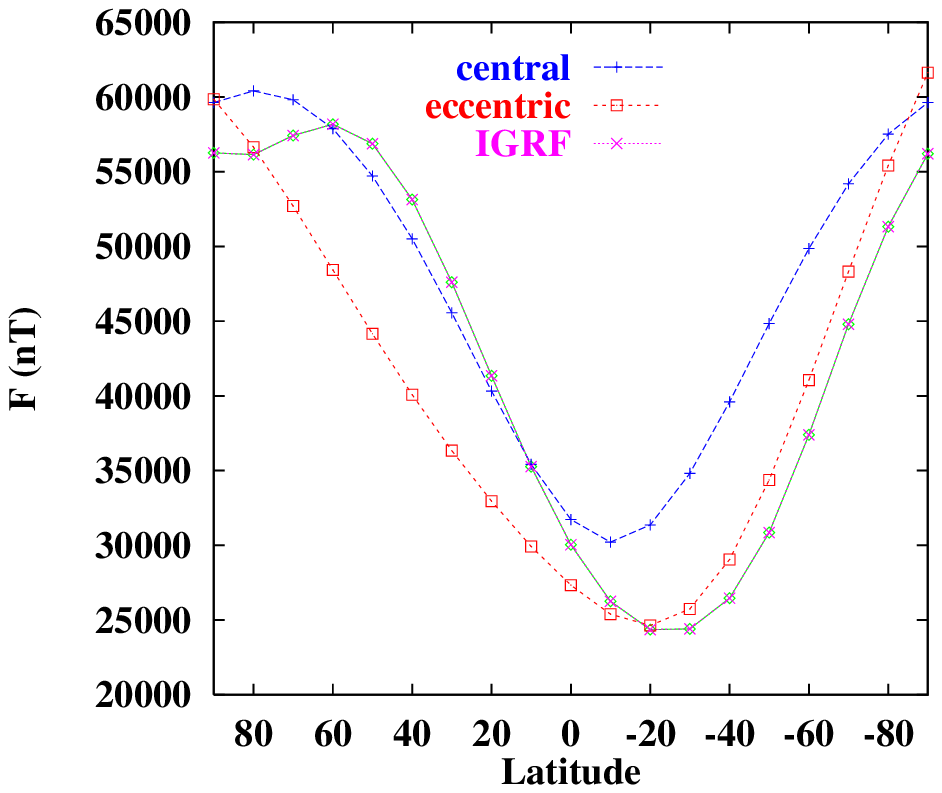,width=7cm}} &
\hbox{\epsfig{file=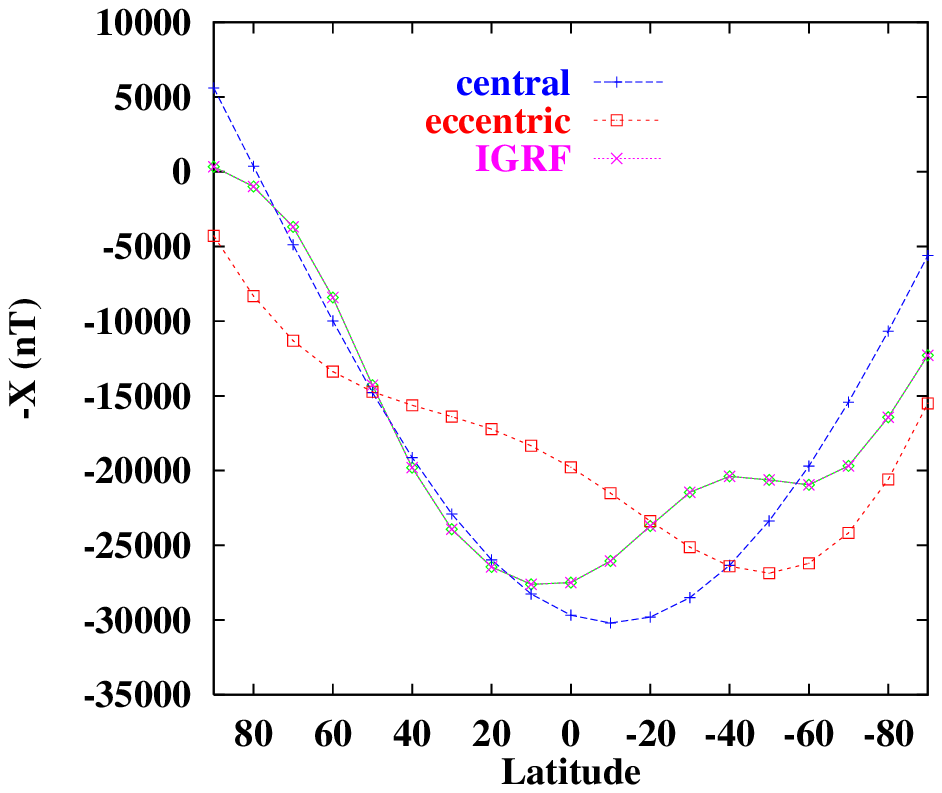,width=7cm}} \\
\hbox{\epsfig{file=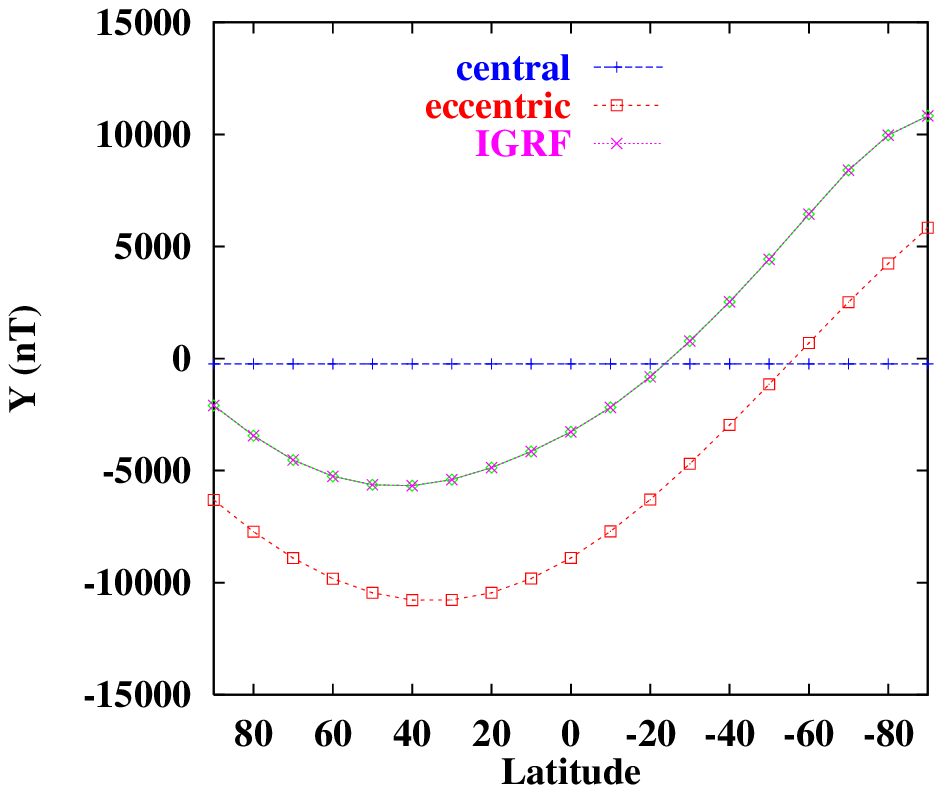,width=7cm}} &
\hbox{\epsfig{file=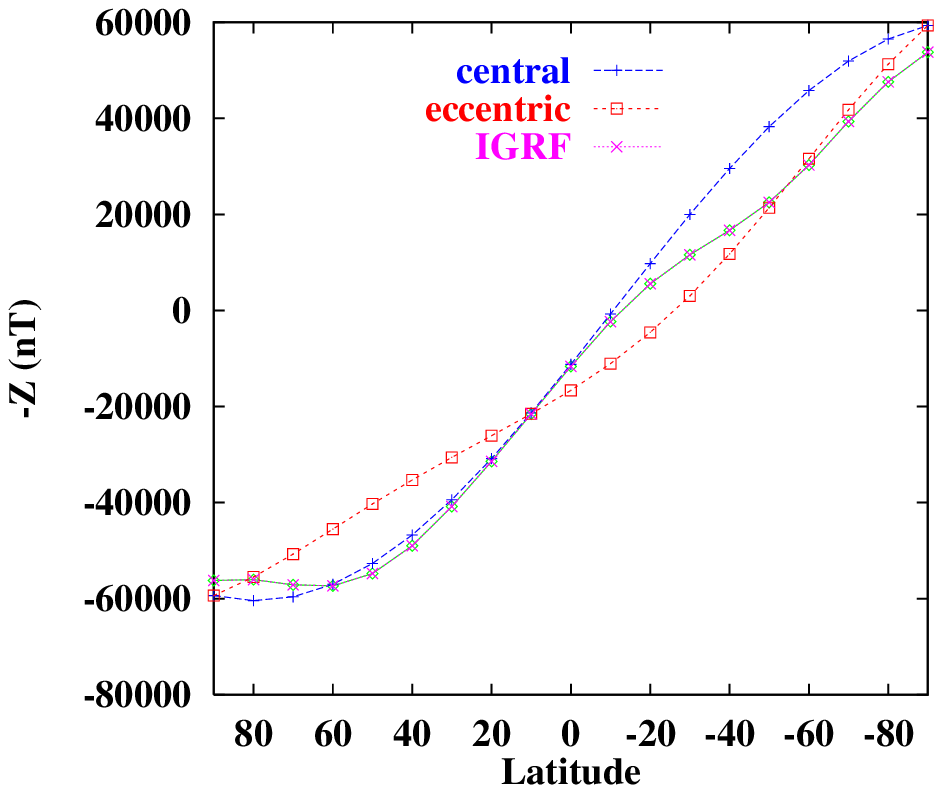,width=7cm}}
\end{array}
\end{displaymath}
\caption{Comparison of different GF models (dipolar centered and
eccentric, and IGRF models).The Cartesian components of the GF are
plotted versus geographic latitude (fixed longitude: $69^\circ$ W).}
\end{figure}

The facts presented so far allow to establish the IGRF as the most
convenient model that can give accurate estimations of the GF to be used in
air-shower simulations.
As a final test, we have checked the IGRF predictions against experimental
data \cite{acacias}.

\begin{figure}
\begin{center}
\epsfig{file=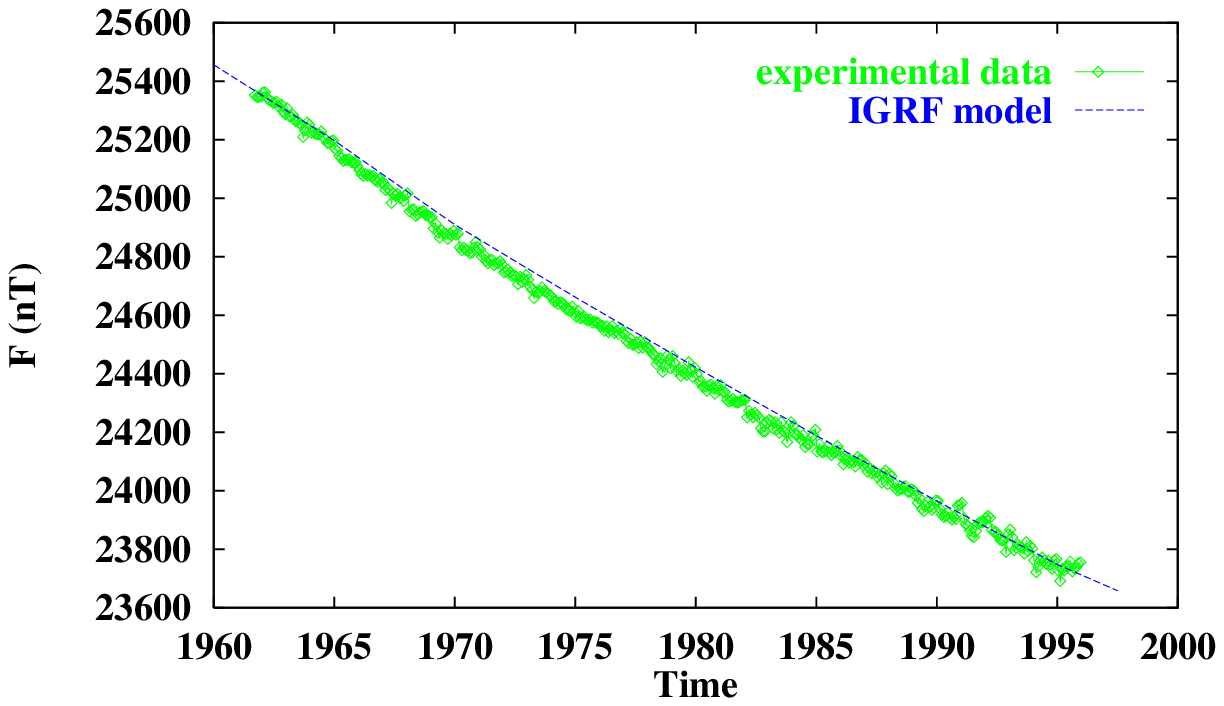,width=11cm}\\
\epsfig{file=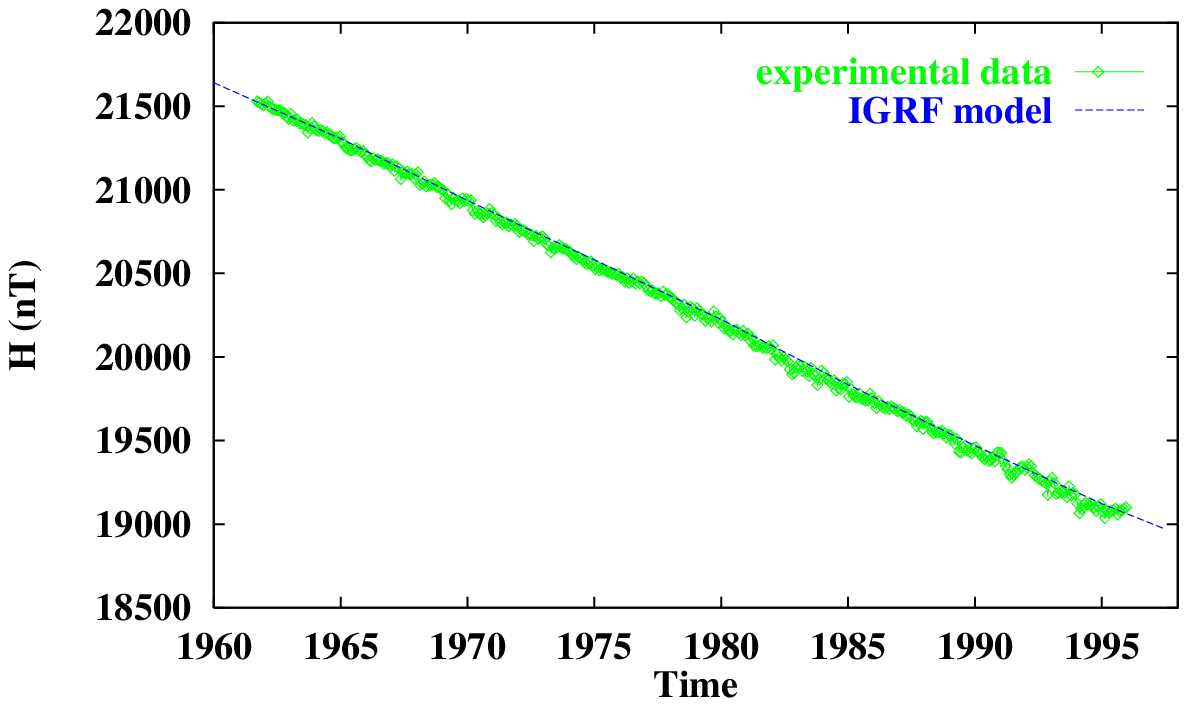,width=11cm}\\
\epsfig{file=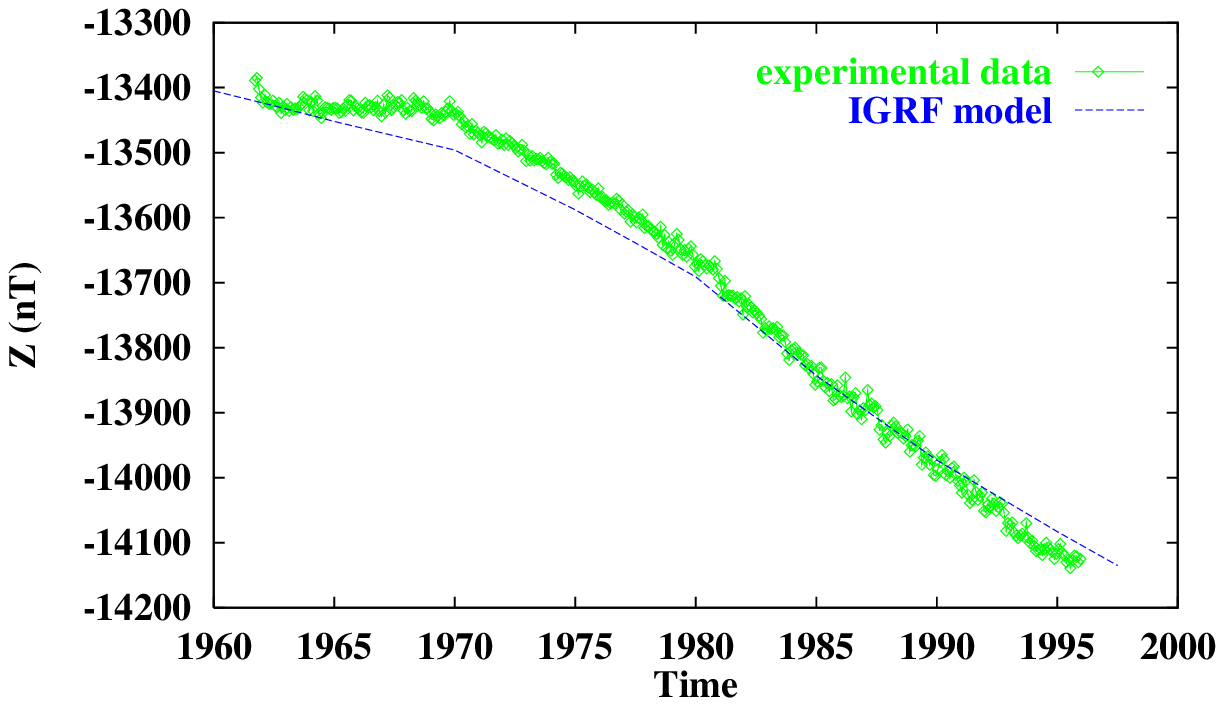,width=11cm}
\end{center}
\caption{ Comparison between experimental data and the IGRF model. The
absolute difference between the experimental and the IGRF prediction is
always less than 200 nT.(Experimental data: Las Acacias observatory,
Argentina \cite{acacias}).}
\end{figure}
In figure 3 the F, H and Z components are plotted against time. The absolute
difference between estimated and measured fields is always less than 200 nT
which is below the 500 nT error bound previously mentioned.

The accuracy of the IGRF predictions is also maintained during magnetic
storms. In figure 4 we illustrate this fact. The experimental data were
registered at the Trelew observatory in Argentina and corresponds to a
magnetic storm that took place during 1994. Again we can see that the 500 nT
bound is always larger than the actual errors. The analysis of the errors in
the declination and inclination angles (not displayed here) indicates that
such errors are always less than 0.5 degree.
\begin{figure}
\begin{center}
\epsfig{file=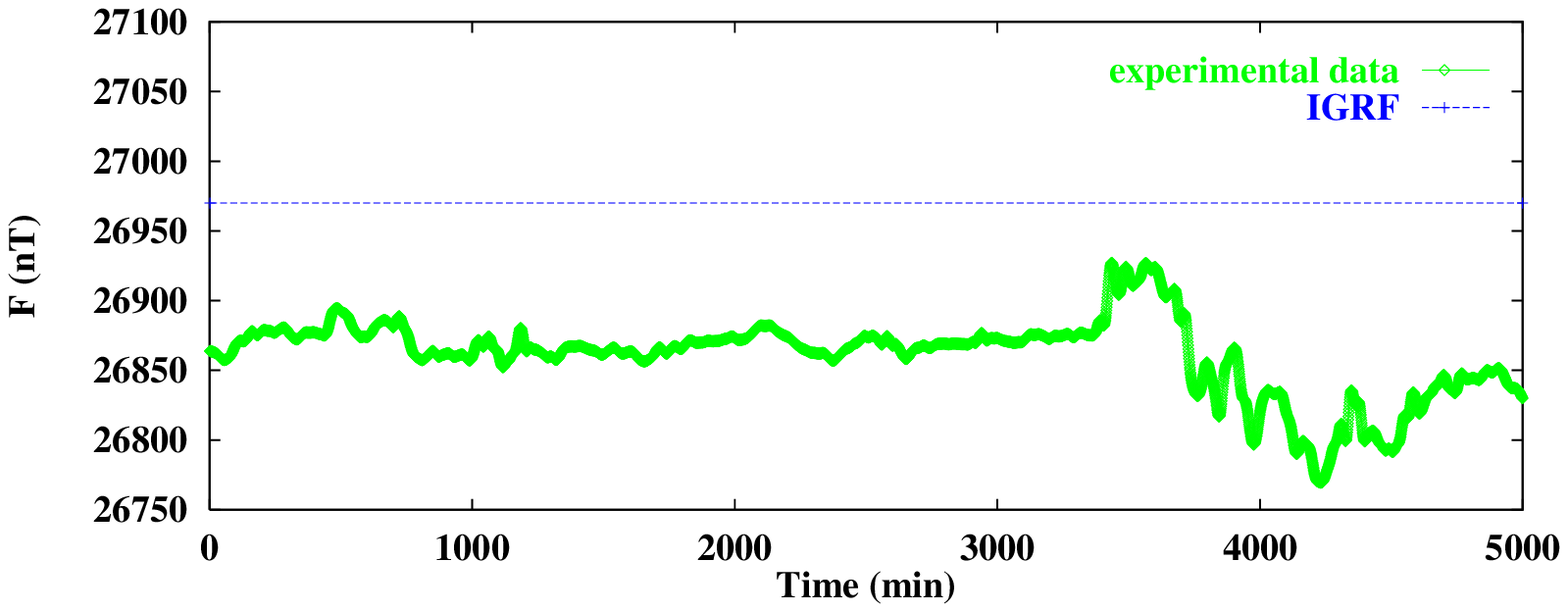,width=12cm}\\
\epsfig{file=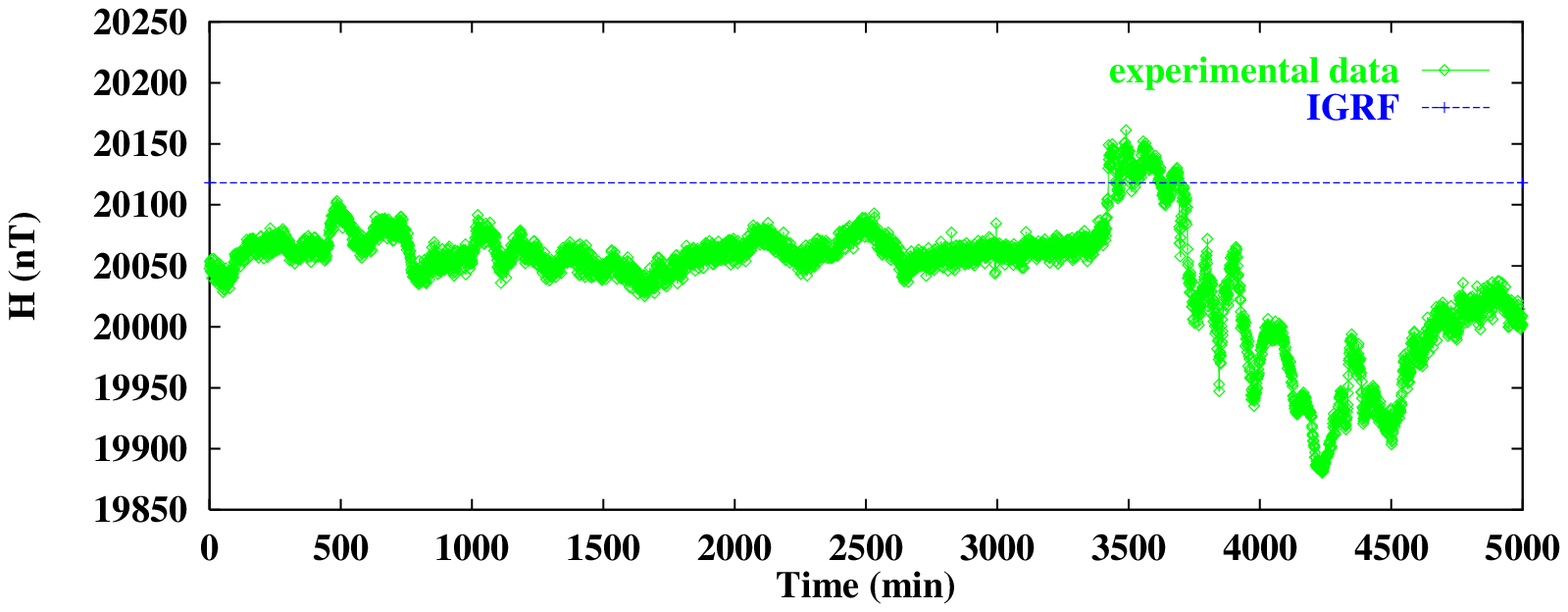,width=12cm}
\end{center}
\caption{ Magnetic storm: Experimental data and IGRF model. The difference
between the experimental and the IGRF prediction is always less than 500 nT.
(Experimental data: Trelew observatory, Argentina \cite{acacias}).}
\end{figure}

\section{Practical Implementation}

The calculation of the GF in any place and time as well as the dynamics of
charged particles in such field have been incorporated into the AIRES program 
\cite{sergio}.

\subsection{Calculation of the GF in the AIRES program}

Special subroutines using the IGRF model for the calculation of the
geomagnetic field have been incorporated to the AIRES program
\cite{sergio,IGRF}.

The GF calculations are controlled from the input instructions. By
means of suitable IDL directives \cite{aires,sergio} the user can
either specify a date and the geographic coordinates of a site to
allow automatic IGRF calculations, or enter manually the GF components
(intensity, inclination and declination).

It is assumed that the shower develops under the influence of a constant and
homogeneous magnetic field which is evaluated before starting the
simulations. Since the region where the shower develops is very small when
compared with the Earth's volume, the mentioned approximation of a constant
and homogeneous field is amply justified.

\subsection{Dynamics of charged particles}

Let us consider the motion of a particle with charge $q$ in a uniform, static
magnetic field $\mathbf{B}$ . The equations of motion are (MKS units): 
\begin{equation}
\frac{d\mathbf{p}}{dt}=q\ \mathbf{v}\times \mathbf{B}  \label{motion}
\end{equation}
where $q,$ $\mathbf{v,}$ and $\mathbf{p}$ are the particle's charge,
velocity and linear momentum respectively.

Since the particle's energy is constant in time, the magnitude of its
velocity is constant and so the Lorentz factor, $\gamma$. The equation of
the unit velocity vector, $\widehat{\mathbf{u}}=\frac{\mathbf{v}}{\shortmid 
\mathbf{v\shortmid }}$, can then be written 
\begin{equation}
\frac{d\widehat{\mathbf{u}}}{dt}=\QOVERD( ) {qc^2}{E}\widehat{\mathbf{u}}%
\times \mathbf{B}  \label{umotion}
\end{equation}
where $E$ is the total energy of the particle (rest plus kinetic) and
$c$ is the speed of light.

As it is well known, the trajectory of a charged particle interacting with a
uniform magnetic field is an helix whose axis is parallel to $\mathbf{B.}$
The motion in a plane normal to the magnetic field is circular, with angular
velocity
\begin{equation}
\omega =\frac{qc^2B}E.  \label{w}
\end{equation}

If the particle advances a distance $\Delta s$ in a time $\Delta t$ ($\Delta
s=\beta c\Delta t$, $\beta =v/c$), the motion can be approximately
calculated via:

\begin{equation}
\widehat{\mathbf{u}}(t+\Delta t)\cong \widehat{\mathbf{u}}(t)+
\frac{d\widehat{\mathbf{u}}}{dt}%
\Delta t=\widehat{\mathbf{u}}(t)+\QOVERD( ) {qc^2\Delta t}{E}
\widehat{\mathbf{u}}%
\times \mathbf{B} .  \label{u}
\end{equation}

For this approximation to be valid, it is needed that 
\begin{equation}
\omega \Delta t=\frac{\omega \Delta s}\beta \ll 1  \label{w2} .
\end{equation}

We have studied the deflection of particles in various representative cases
finding that equation (\ref{w2}) is always satisfied for all the particles
that are tracked during the simulation of air-showers, even in the least
favorable case of low energy electrons or positrons.
It is therefore safe to use equation (\ref{u}) to account for the
deflections of charged particles moving under the effect of the GF.

The magnetic deflection algorithm implemented in AIRES makes use of equation
(\ref{u}) to evaluate the updated direction of motion at time $t+\Delta t$.
However, it also uses a ``technical trick'', inspired in a similar procedure
used in the well-known program MOCCA \cite{mocca}: The path $\Delta s$ is
divided in two halves of length $\Delta s/2$ each. Then the particle is
moved the first half using the old direction of motion
$\widehat{\mathbf{u}}(t)$, and
the second one with the updated vector $\widehat{\mathbf{u}}(t+\Delta
t).$ This means
that the correction at time $\Delta t$ is applied starting at $t=\Delta t/2$
 and, as a result, compensates for the inaccurate direction
$\widehat{\mathbf{u}}(t)$ used in the first half.

\begin{figure}
\begin{center}
\makebox{{\input{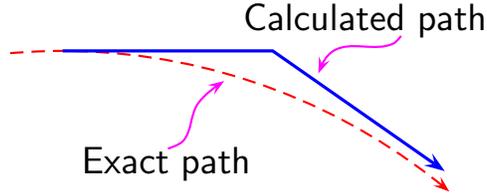}}}
\end{center}
\caption{ Schematic representation illustrating the algorithm used in AIRES
to move charged particles under the influence of the GF. The complete path
is divided in two halves each one of length $\Delta s/2$. The particle is
advanced the first half using the current direction of motion. The direction
of motion is updated to its final value and finally the particle is advanced
the second half using this new direction. In this figure the radius of
curvature was exaggeratedly reduced to 10 cm in order to make visible the
difference between exact and calculated paths. For realistic conditions such
differences are completely negligible.}
\end{figure}
In figure 5 the advancing-deflecting procedure is represented
schematically. Here the radius of curvature of the exact trajectory was
deliberately reduced to make evident the differences between the exact and
approximate paths. In the case of realistic curvature radii (of hundreds or
thousands of meters) both exact and simulated trajectories are virtually
coincident.

\section{Simulations}

We have analyzed the influence of the GF on air shower observables performing
some simulations in a variety of initial conditions.

We have simulated 10$^{19}$ eV proton air showers with varying injection
altitudes and zenith angles (from 0 to 80 deg), thinning levels range from 10%
$^{-4}$ to 10$^{-7}$ relative.
We have chosen the Millard county (Utah, USA) site where the GF is intense
and therefore the differences between the simulations with and without field
can be appreciated better.

The direct inspection of global observables such as shower maximum, total
number of particles at ground, etc.~shows no evident effect of the GF. Some
tendencies can be detected when a large number of showers is simulated,
and in most cases the deviations are of the order of the fluctuations,
either natural or induced by thinning.

However, differences do appear when a detailed analysis of some particle
distributions is made. To illustrate this point we are going to report here
the results of some of our simulations.

All the selected showers were simulated at 10$^{-7}$ relative thinning level
in order to obtain clear data; the magnetic field, when enabled, amounts
to 52800 nT (F) with an inclination ${\rm I}=64.8$ degrees; the zenith
angle is 70
degrees and the azimuth 90 degrees (incidence plane normal to the magnetic
north). The ground level is located at 1000 m.a.s.l (920 g/cm$^2$) and we
have considered two injection altitudes case A: 9350 m ($X=300$ g/cm$^2$) and
case B: 100 km ($X\cong 0$).

The mean positions of the shower maximum are, approximately, 560 g/cm$^2$
(case A) and 280 g/cm$^2$ (case B). The path from the mean shower maximum
down to the ground level, measured along the shower axis, for case A (B) is
1050 g/cm$^2$ (1870 g/cm$^2$). As a result, the attenuation of the
showers at ground level is larger for case B than for case A; and the number
of particles reaching ground is thus much smaller for case B. For example,
without the effect of the GF the total number of electrons and positrons
reaching ground for case A is $(1.6\pm 0.4)\times 10^8$, while the figure
corresponding to case B is $(4.7\pm 0.5)\times10^6$, that is 34 times smaller
than the previous case.

The reason for selecting these two cases was to investigate the influence of
the GF in two different phases of the shower development: (A) shortly after
reaching its maximum and (B) when the shower is about to vanish completely.

\begin{figure}
\begin{center}
\epsfig{file=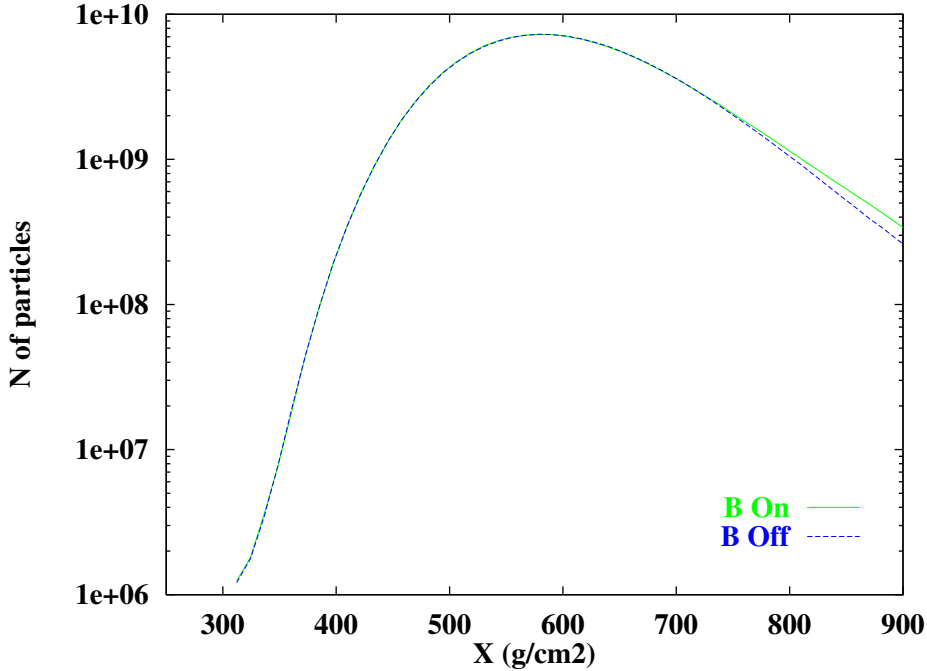}
\end{center}
\caption{ Longitudinal development of all charged particles (case A).}
\end{figure}
Let us start our comparative analysis studying the longitudinal
development of all charged particles. In figure 6, the total number of
charged particles is plotted against the vertical depth $X$. The
initial conditions are the ones corresponding to the case A. Comparing
the plots coming from the simulations with and without GF we can see
that there is no significant difference either in the position of the
maximum ($X_{\rm max}$) or in the maximum number of charged particles
($N_{\rm max}$). Even if the showers simulated without GF seem to
vanish more quickly as $X$ grows, this difference may not be
significant since it is never larger than one standard deviation of
the corresponding mean.

\begin{figure}
\begin{center}
\epsfig{file=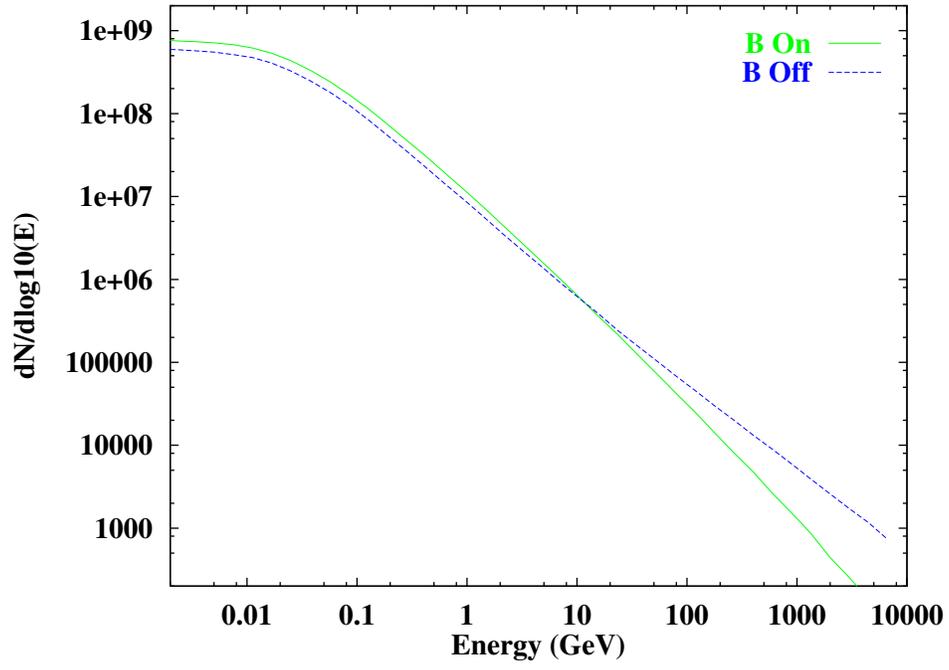}
\end{center}
\caption{$\gamma $ ground energy distribution (case A).}
\end{figure}
\begin{figure}
\begin{center}
\epsfig{file=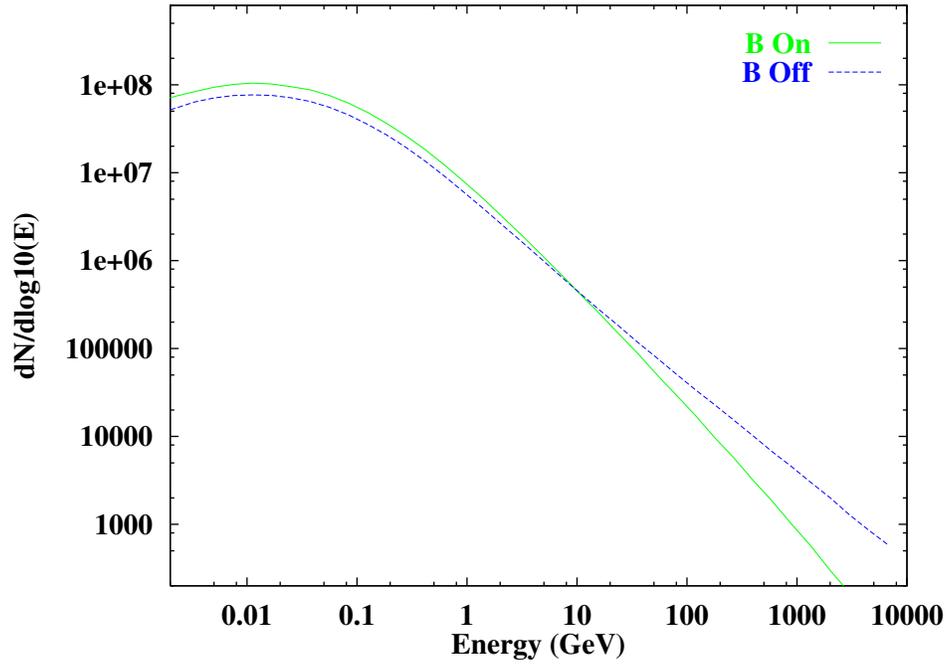}
\end{center}
\caption{$e^{+}$ and $e^{-}$ ground energy distribution (case A).}
\end{figure}
The ground level energy distributions of $\gamma $'s and $e^{\pm }$'s (case
A) are plotted in figures 7 and 8 respectively. When the GF is enabled both
ground energy distributions change in the same way: In both cases the number
of particles that reach the ground increases for low energy particles and
decreases for high energy ones. The correlations between the respective
distributions can be understood taking into account that ground electrons,
positrons or gammas surely come from the ``nearby'' electromagnetic cascade
whose intrinsic mechanisms constantly generate gammas from $e^{\pm }$ and
vice-versa, being the energies of the secondaries lower (at most equal) than
those of the respective primaries. It is therefore clear that the structure
of both energy distributions should be similar.
When the GF is considered, the trajectory of electrons and positrons are
helicoidal.
This generally leads to an increase of the particles's paths and
therefore to
larger continuum energy losses (ionization losses), diminishing (enlarging)
the average number of high (low) energy particles that reach ground.

The ground $\mu ^{\pm }$ energy distributions (case A) are plotted in figure
9. There are no measurable differences between the energy distributions
corresponding to the cases with and without GF for these particles. The
muons travel long distances without interacting and their radii of
curvature are generally larger than for $e^{\pm }.$ This implies that these
particles reach the ground without important increases in their paths and
thus without modifications in their energy.\footnote{%
In the current AIRES version the energy of the muons is altered by continuum
loss mechanisms and/or emission of knock-on electrons. Muon bremsstrahlung
is not taken into account yet.}
\begin{figure}
\begin{center}
\epsfig{file=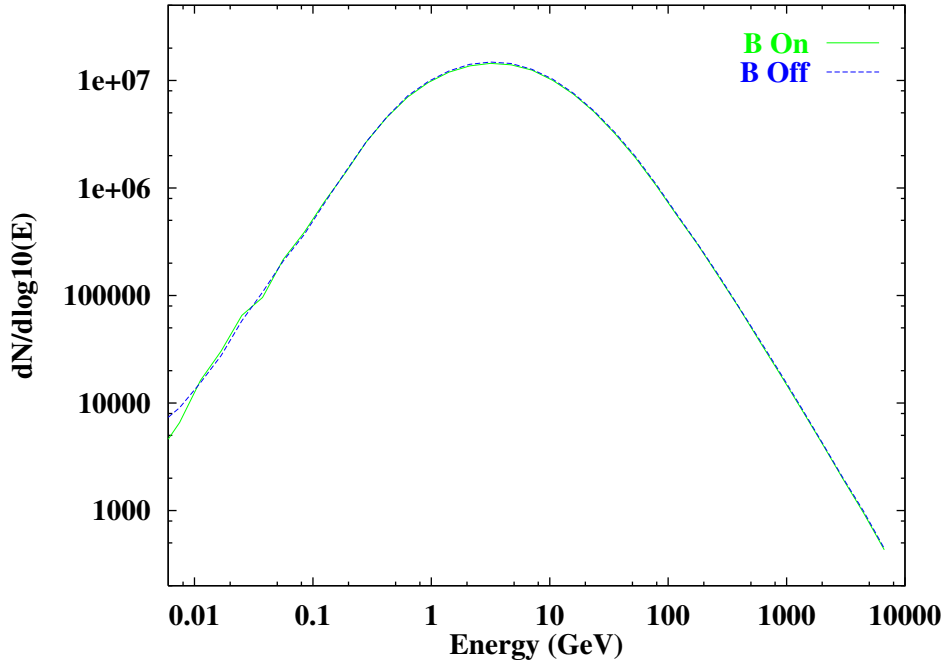}
\end{center}
\caption{$\mu ^{+}$ and $\mu ^{-}$ ground energy distribution (case A).}
\end{figure}

On the other hand, the lateral distribution of ground muons does present
significant modifications when the GF is considered.

To evaluate the densities plotted in figures 10 (case A) and 11 (case B) we have
filtered all the muons arriving at ground at points $(x_g,y_g)$ verifying 
$\left|\arctan\frac{y_g}{x_g}\right|<10^\circ$. In other words, we
have selected
particles lying in the region close to the $x$-axis (in polar
coordinates, no more
than 10 degrees apart) where the effects of the magnetic deflection are most
noticeable. To make the corresponding histograms, both the positive and
negative $x$-axis were divided in radial bins $[r_i,r_{i+1}]$, with
$r_{1\ldots8}=150$, 200, 300, 400, 600, 800, 1200 and 1600 meters.

The symmetry of distributions around the origin when the GF is disabled
shows up clearly in figures 10 and 11 where the $\mu ^{+}$ and $\mu ^{-}$
distributions (${\bf B}$ off) are virtually coincident.

\begin{figure}
\begin{center}
\epsfig{file=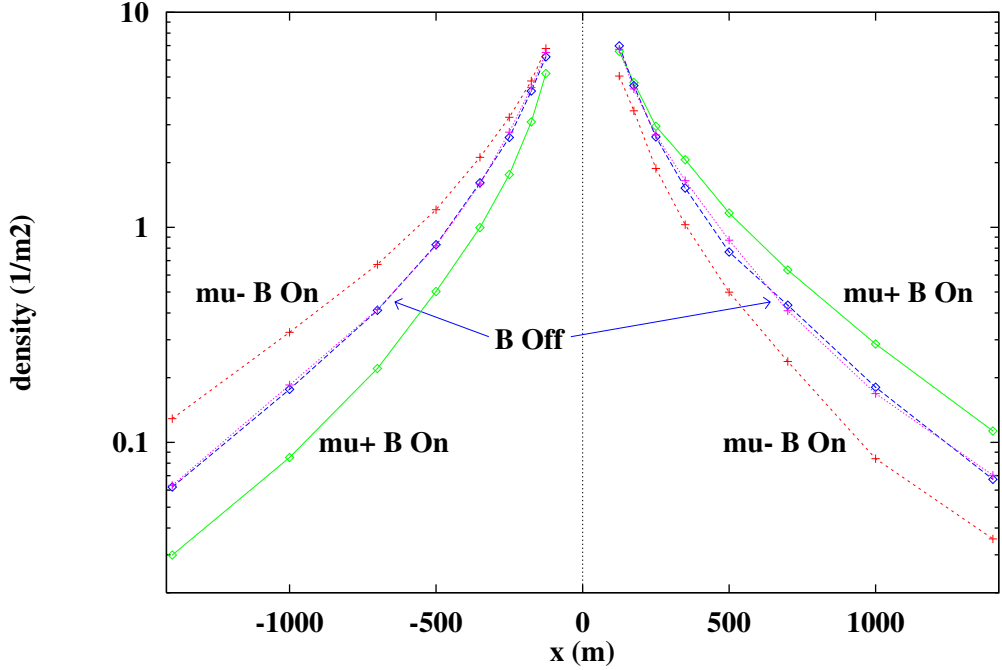}
\end{center}
\caption{ Density of $\mu ^{+}$ and $\mu ^{-}$ for particles arriving near
the $x$-axis, case A (see text).}
\end{figure}
\begin{figure}
\begin{center}
\epsfig{file=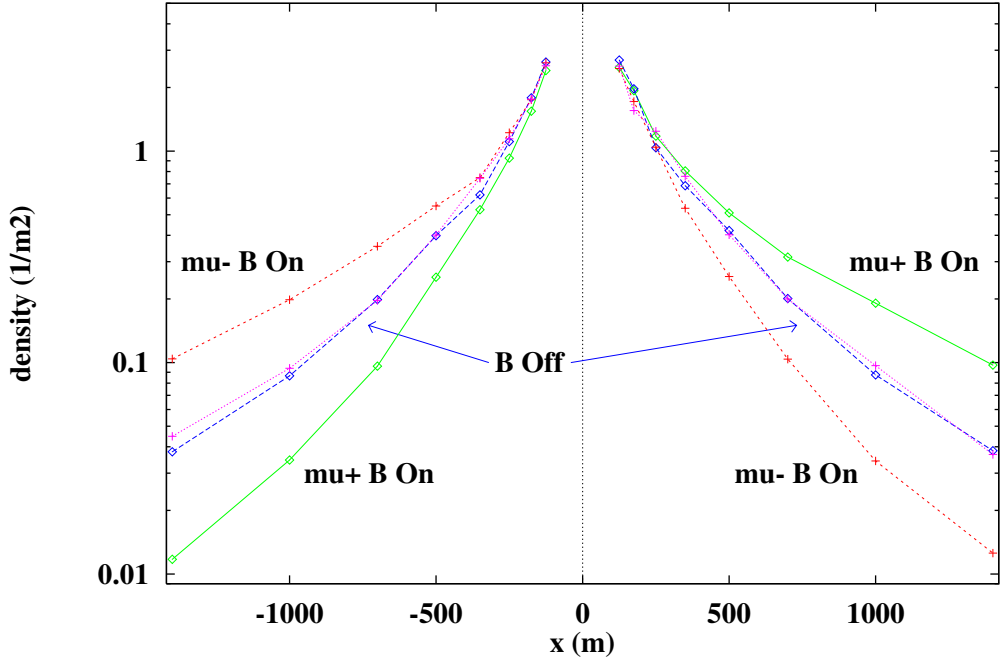}
\end{center}
\caption{ Same as figure 10, but for case B.}
\end{figure}
On the other hand, when the GF is on, a displacement of the density of
positive (negative) particles towards positive (negative) $x$-axis occurs.
Fitting the simulation data to suitable distribution functions with the
central position as a free parameter, it is possible to estimate the
difference between the peaks in the $\mu ^{+}$, $\mu ^{-}$ distributions in
about 70 m.

The difference between the numbers of $\mu ^{+}$ and $\mu ^{-}$ increases
with the distance to the shower core arriving to about one order of
magnitude at $|x|=2000$ m for the case of
completely developed showers (figure 11). Notice also that the total number
of muons ($\mu ^{+}$ and $\mu ^{-}$) also changes when the GF is switched on.

Another way of studying the influence of the GF on the muon distribution is
to analyze the dependence of the ground particle density $\rho (r,\theta )$
with the polar angle $\theta $, for $r$ belonging to a certain interval $%
[r_1,r_2].$

The $\mu ^{\pm }$ densities versus polar angle $\theta $ are plotted in figures
12 (case A) and 13 (case B), for $[r_1,r_2]=[100\>{\rm
m},200\>{\rm m}]$ and [300 m, 600 m] .
\begin{figure}
\begin{displaymath}
\begin{array}{cc}
\hbox{\epsfig{file=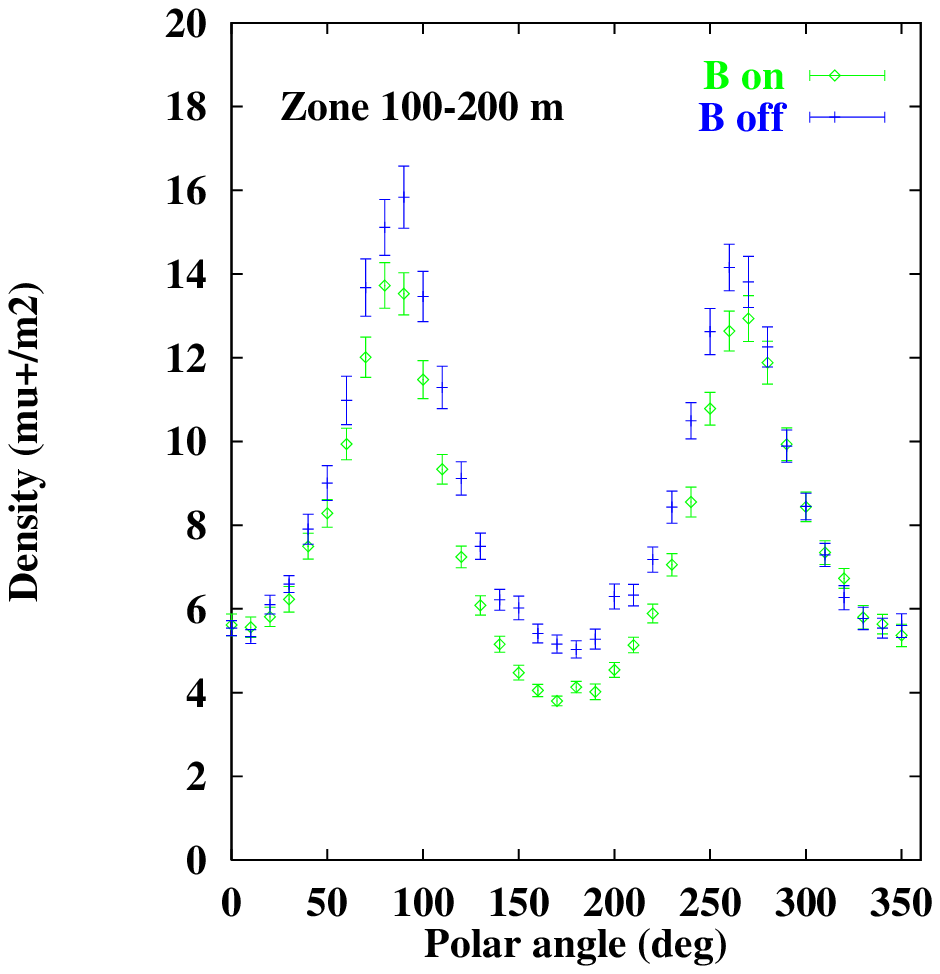,width=7.25cm}} &
\hbox{\epsfig{file=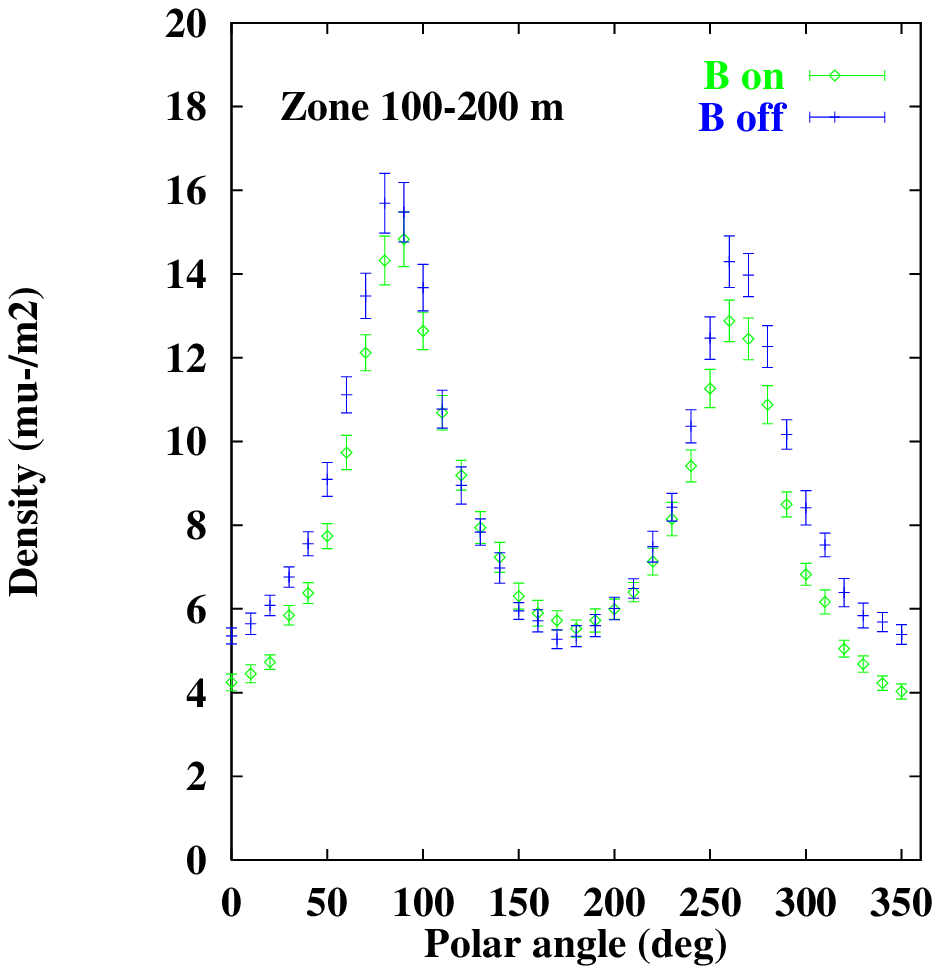,width=7.25cm}} \\
\hbox{\epsfig{file=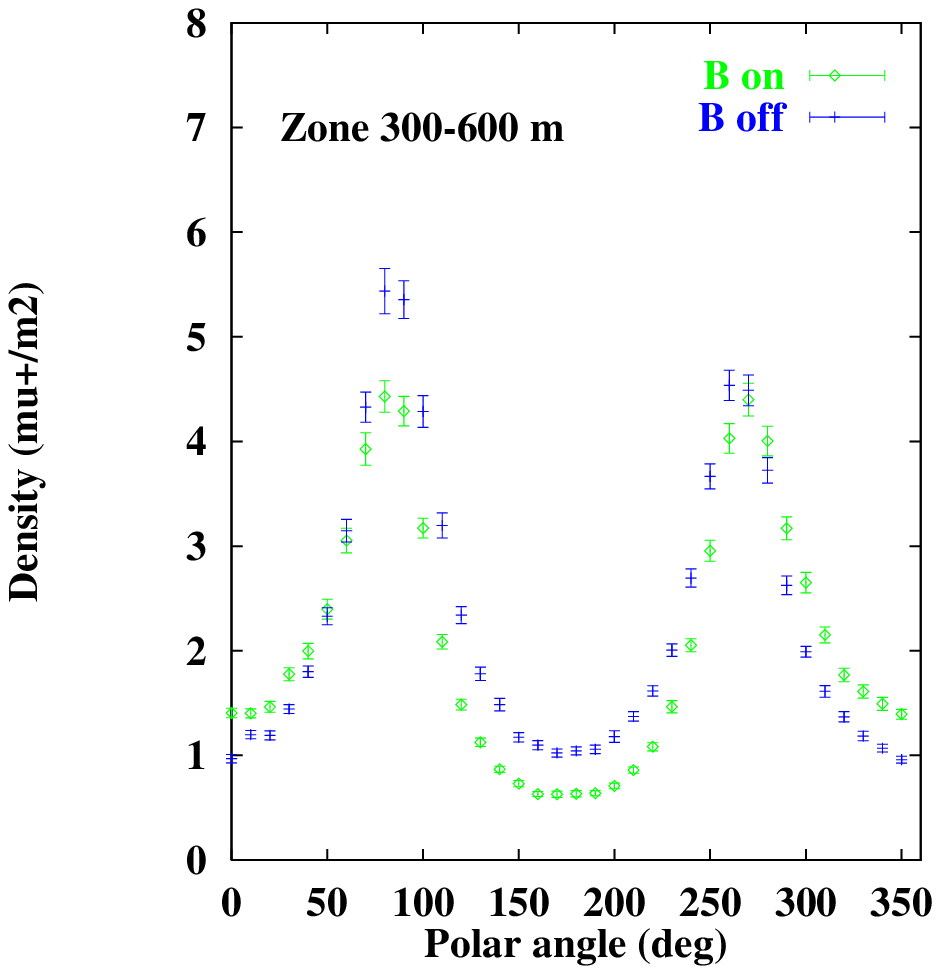,width=7.25cm}} &
\hbox{\epsfig{file=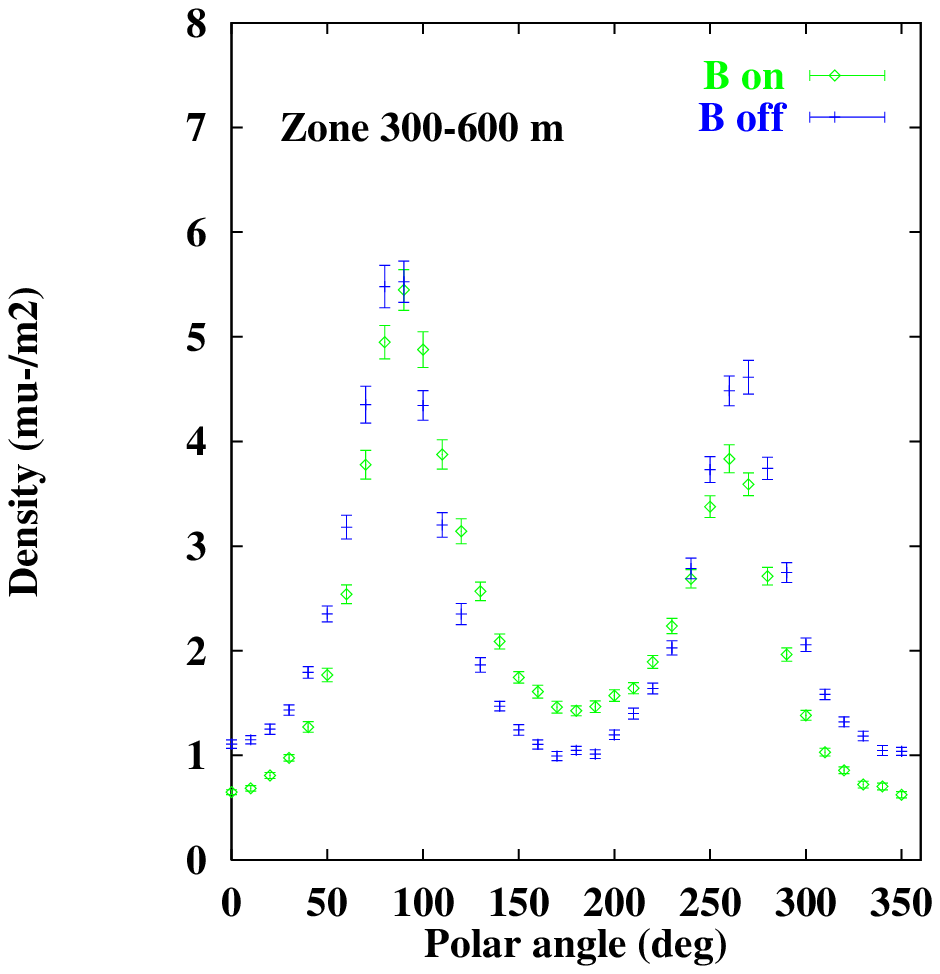,width=7.25cm}}
\end{array}
\end{displaymath}
\caption{ Density of $\mu ^{+}$ and $\mu ^{-}$ at ground $\rho (r,\theta )$
versus polar angle $\theta$ for $100 \le r\le 200$ m and $300 \le r \le
600$ m (case A).}
\end{figure}
\begin{figure}
\begin{displaymath}
\begin{array}{cc}
\hbox{\epsfig{file=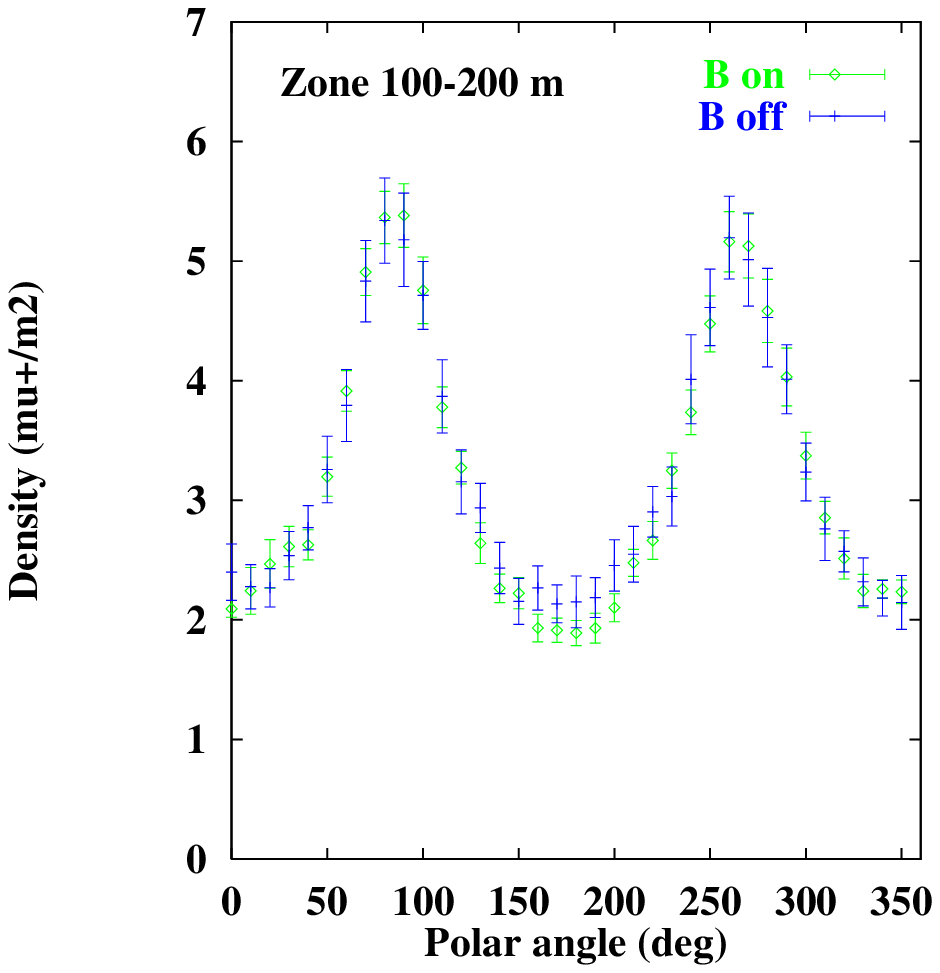,width=7.25cm}} &
\hbox{\epsfig{file=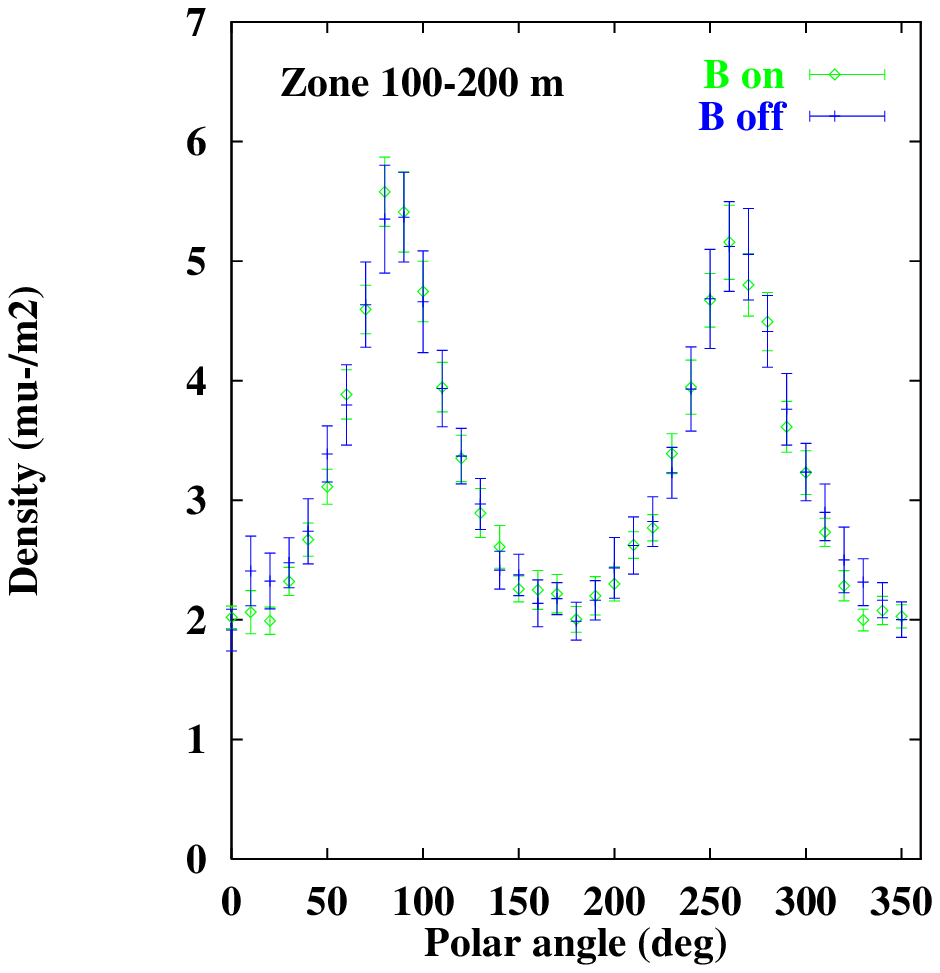,width=7.25cm}} \\
\hbox{\epsfig{file=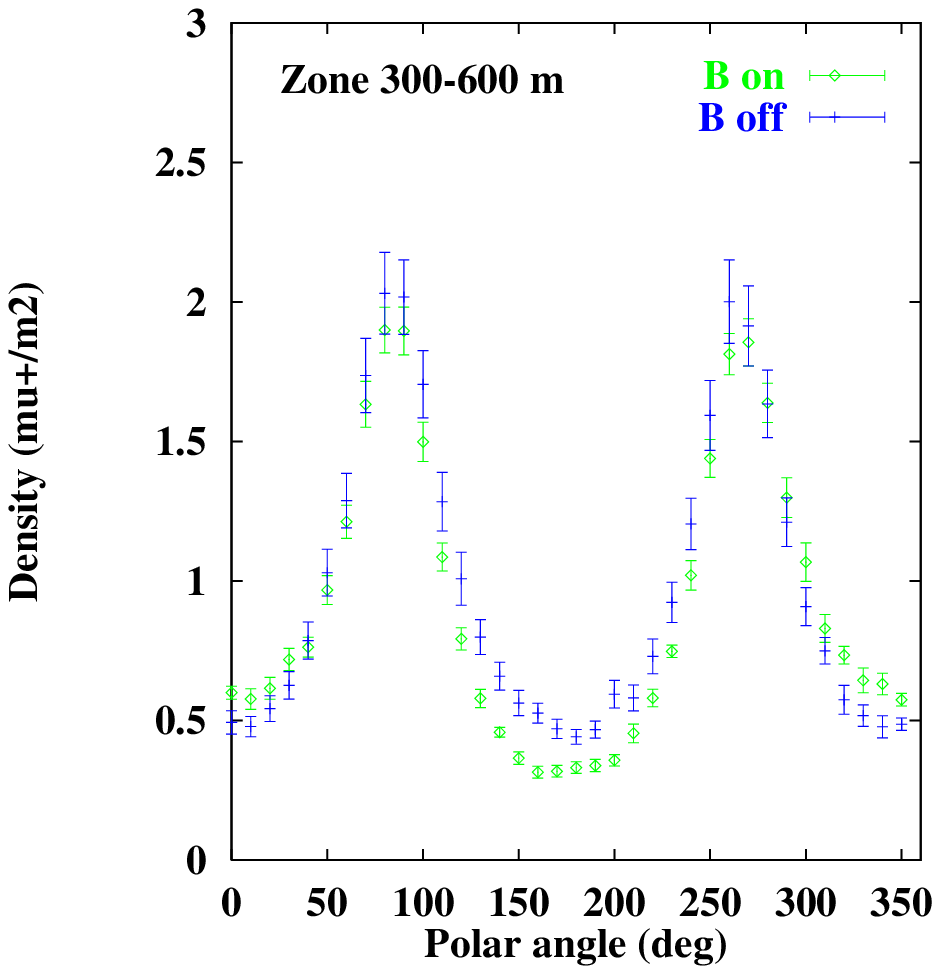,width=7.25cm}} &
\hbox{\epsfig{file=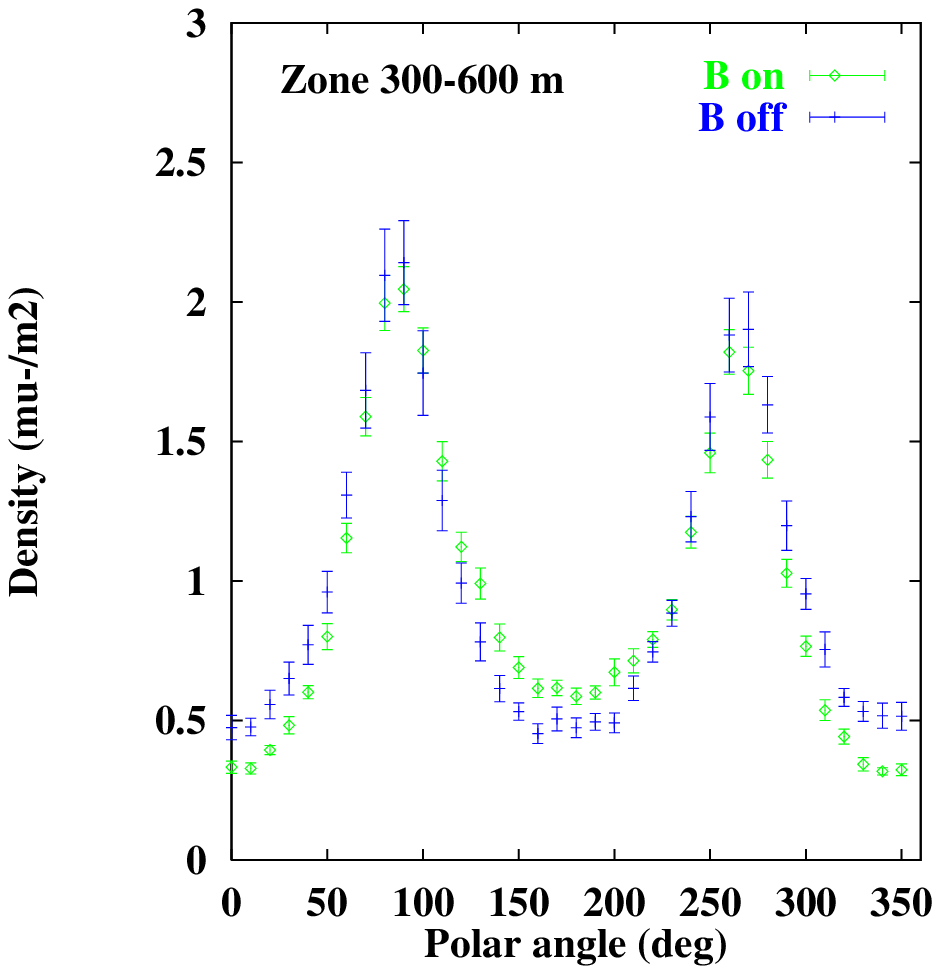,width=7.25cm}}
\end{array}
\end{displaymath}
\caption{ Same as figure 12, but for case B.}
\end{figure}

When the GF is off, the plotted data show the anisotropy due to the nonzero
zenith angle of the shower axis. Notice that all the densities reach their
maximum in the zone that is closest to the arrival direction ($\theta \cong
90$ deg).

The plots corresponding to the simulations with the GF enabled are
consistent with the data displayed in figures 10 and 11: The $\mu ^{+}$ ($\mu
^{-})$ distribution is larger in the positive (negative) $x$-axis region, $%
\theta \cong 0^\circ$ ($\theta \cong 180^\circ$).

Such characteristics show up clearly in the plots corresponding to the
[300 m, 600 m] zone. For the [100 m, 200 m] zone the approximately unaltered
number of $\mu ^{+}$ ($\mu ^{-})$ in the proximity of $\theta =0^\circ$ ($%
\theta =180^\circ$) is due to the fact that the zone is located too much near
the peak of the distribution (See figure 10).

In the case of electrons and positrons, the fluctuations are large enough to
prevent the detection of any relevant difference in particle distributions.
This clearly shows up in figure 14 where the $e^{\pm }$ densities (case A)
were plotted in similar conditions as in the $\mu ^{\pm }$ case.
\begin{figure}
\begin{displaymath}
\begin{array}{cc}
\hbox{\epsfig{file=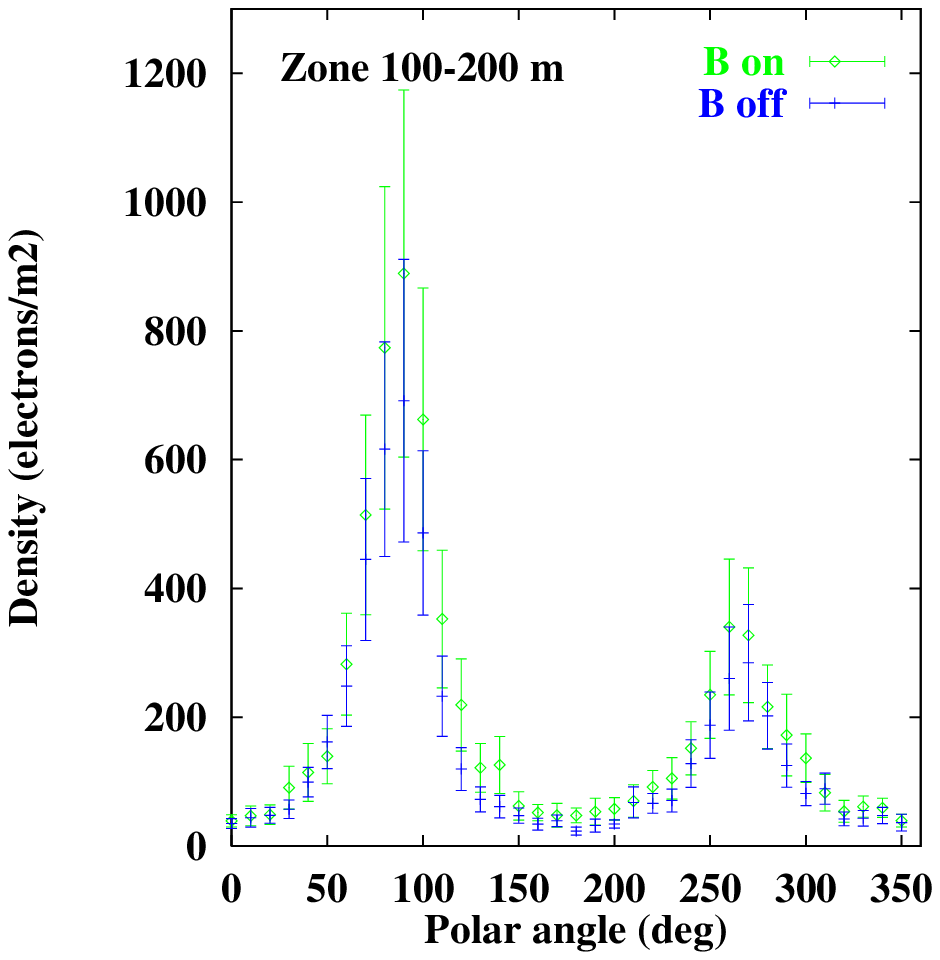,width=7.25cm}} &
\hbox{\epsfig{file=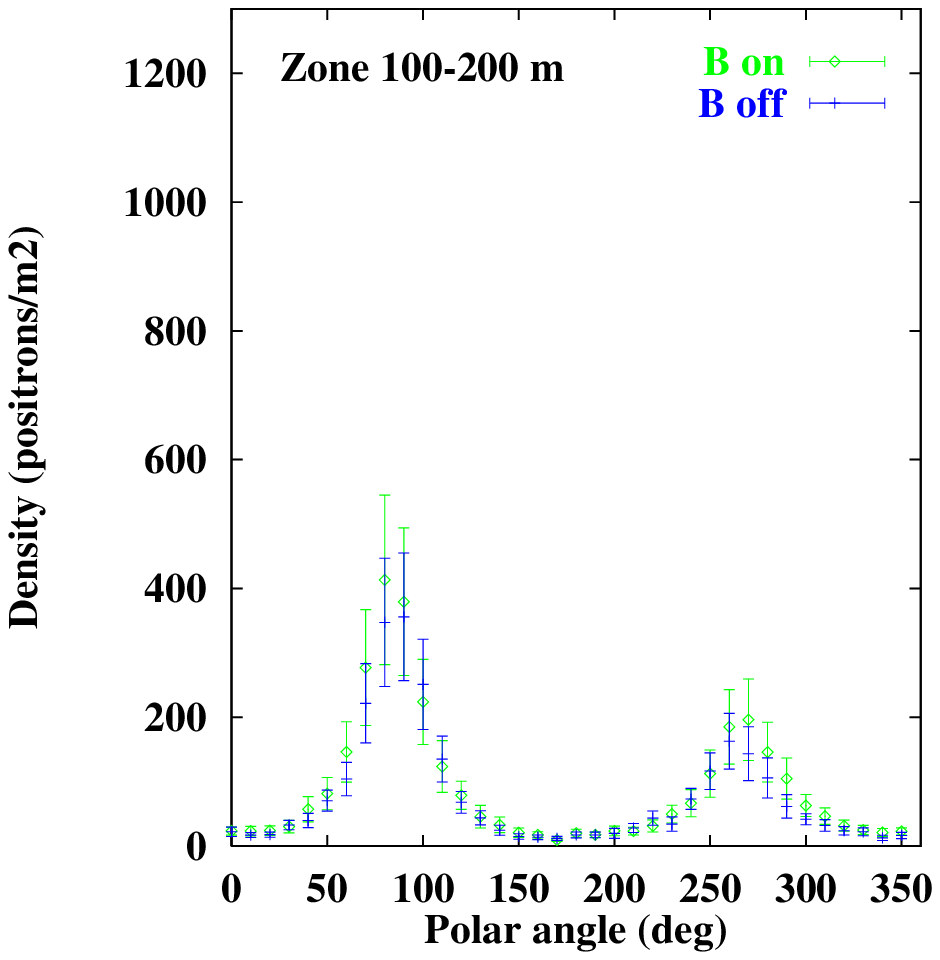,width=7.25cm}} \\
\hbox{\epsfig{file=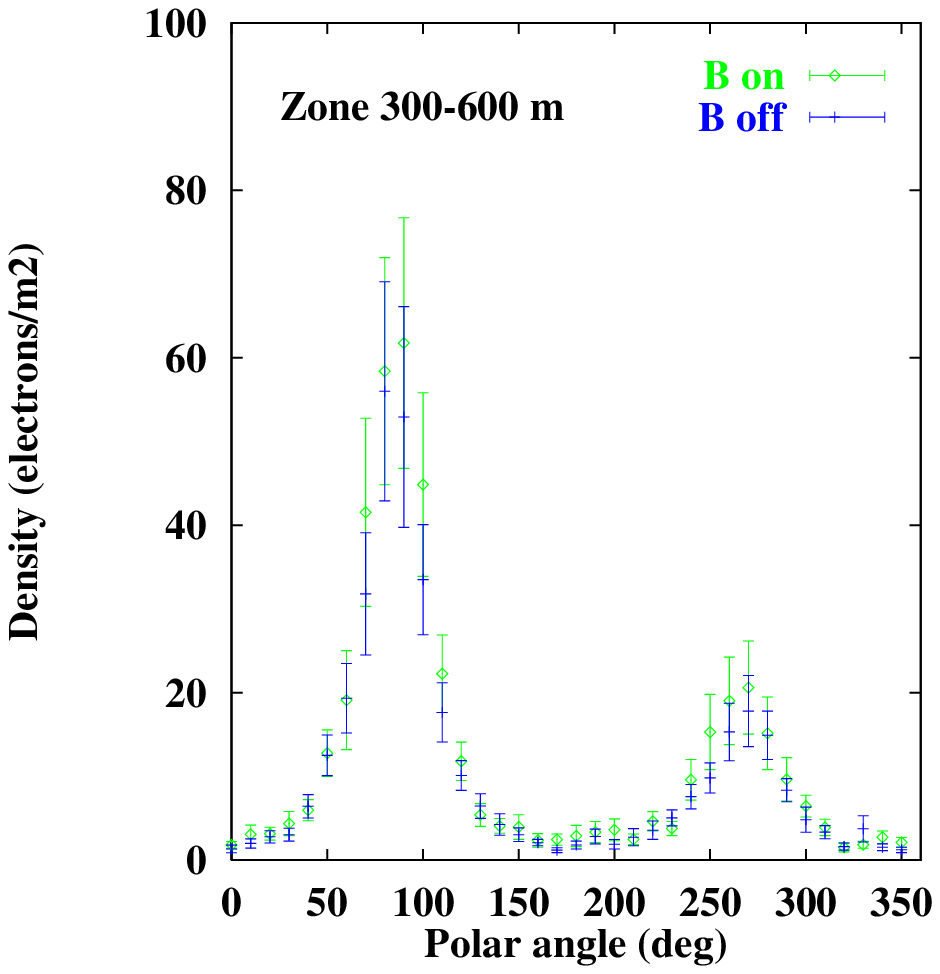,width=7.25cm}} &
\hbox{\epsfig{file=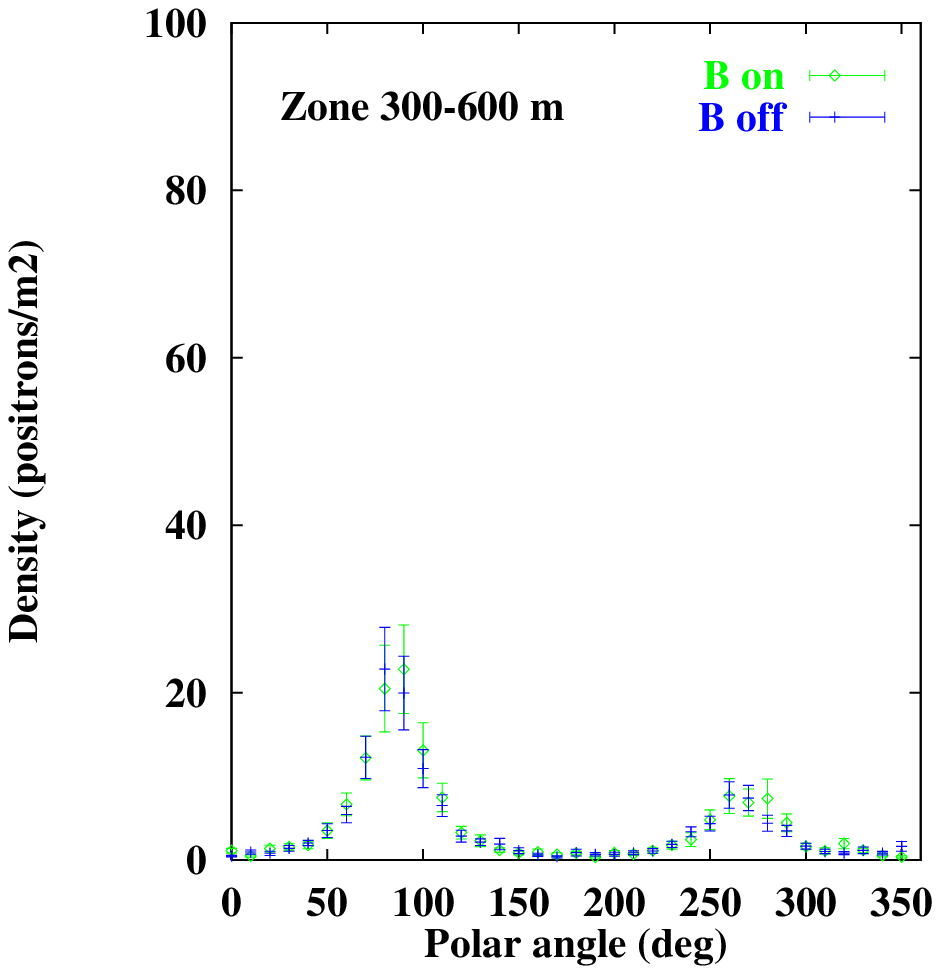,width=7.25cm}}
\end{array}
\end{displaymath}
\caption{ Same as figure 12, but for $e^{+}$ and $e^{-}$ densities (case A).}
\end{figure}

\section{Conclusions}

We have discussed in this work the influence of the geomagnetic field on the
most common observables that characterize the air showers initiated by
astroparticles. The data used in our analysis were obtained from computer
simulations performed with the AIRES program.

Our work includes the analysis of the main properties of the geomagnetic
field, as well as the implementation of the related algorithms in the
program AIRES.

By means of the International Geomagnetic Reference Field (IGRF) it is
possible to make accurate evaluations of the average geomagnetic field at
a certain place given its geographical coordinates, altitude above sea
level and time.

We have used this tool to run the simulations using a realistic geomagnetic
field, selecting the location of Millard county (Utah, USA) as a convenient
place with a high magnetic field intensity.

The changes that the analyzed magnitudes experiment when the
geomagnetic field is taken 
into account are generally small, but there are certain observables like the
lateral distribution of muons where such differences become significant.

Considering these facts we conclude saying that the geomagnetic field
should be taken into account whenever a particular event with precisely
determined initial conditions must be simulated accurately.

In future works we will address another effects on the air shower
development that are related to the geomagnetic field.

\section*{Acknowledgments}

We are indebted to L. N. Epele, C. A. Garc\'{\i}a Canal, and H. Fanchiotti for
useful discussions; also to O. Medina Tanco (S\~{a}o Paulo University,
Brazil) and J. Valdez (UNAM, Mexico) for their help to obtain information
about the IGRF.

The experimental data from Las Acacias and Trelew Observatories are courtesy
of J. Gianibelli (FCAGLP, La Plata, Argentina).

Finally we want to thank C. Hojvat (Fermilab, USA) and C. Pryke (U. of Chicago,
USA) who gave us the possibility of running our simulations on very powerful
machines.

\end{document}